\documentclass[paper]{JHEP3} % 10pt is ignored!
\usepackage{amsmath,amssymb,setspace,graphicx,subfigure,epsfig}

\JHEPspecialurl{http://jhep.sissa.it/JOURNAL/JHEP3.tar.gz}

\newcommand\fverbdo{\egroup\medskip\noindent%
            \fbox{\unhbox\fverbbox}\ }
\newcommand\fverbit{\egroup\item[\fbox{\unhbox\fverbbox}]}
\newbox\fverbbox

\newcommand{\be}{\begin{equation}}
\newcommand{\ee}{\end{equation}}
\newcommand{\bea}{\begin{eqnarray}}
\newcommand{\eea}{\end{eqnarray}}

\newcommand{\eq}[1]{Eq.~(\ref{#1})}
\newcommand{\eqs}[2]{Eqs.~(\ref{#1}) and (\ref{#2})}

\newcommand{\eV}{\mathinner{\mathrm{eV}}}
\newcommand{\keV}{\mathinner{\mathrm{keV}}}
\newcommand{\MeV}{\mathinner{\mathrm{MeV}}}
\newcommand{\GeV}{\mathinner{\mathrm{GeV}}}
\newcommand{\TeV}{\mathinner{\mathrm{TeV}}}

\newcommand{\planck}{\mathrm{P}}

\def\bea{\begin{eqnarray}}
\def\eea{\end{eqnarray}}
\def\be{\begin{equation}}
\def\ee{\end{equation}}

\title{The $\mu$ problem and sneutrino inflation}

\author{Yeong Gyun Kim \\
Gwangju National University of Education, Gwangju 500-703, Korea. \\
E-mail: \email{ygkim@gnue.ac.kr}
}

\author{Hyun Min Lee \\
CERN, Theory division, Geveva 23, CH-1211, Switzerland. \\
E-mail: \email{hyun.min.lee@cern.ch}}

\author{Wan-Il Park \\
School of Physics, KIAS, Seoul 130-722, Korea. \\
E-mail: \email{wipark@kias.re.kr}
}

\abstract{
We consider sneutrino inflation and post-inflation cosmology in the singlet extension of the MSSM with approximate Peccei-Quinn(PQ) symmetry, assuming that supersymmetry breaking is mediated by gauge interaction.  The PQ symmetry is broken by the intermediate-scale VEVs of two flaton fields, which are determined by the interplay between radiative flaton soft masses and higher order terms. Then, from the flaton VEVs, we obtain the correct $\mu$ term and the right-handed(RH) neutrino masses for see-saw mechanism. We show that the RH sneutrino with non-minimal gravity coupling drives inflation, thanks to the same flaton coupling giving rise to the RH neutrino mass. After inflation, extra vector-like states, that are responsible for the radiative breaking of the PQ symmetry, results in thermal inflation with the flaton field, solving the gravitino problem caused by high reheating temperature. Our model predicts the spectral index to be $n_s\simeq 0.96$ due to the additional efoldings from thermal inflation. We show that a right dark matter abundance comes from the gravitino of $100 \,{\rm keV}$ mass and a successful baryogenesis is possible via Affleck-Dine leptogenesis.
}

\begin{document}

%\maketitle  IS IGNORED %%%%%%%%%%%

\section{Introduction}

%mu problem and PQ symmetry

In the Minimal Supersymmetric Standard Model(MSSM), the $\mu$ term is a supersymmetric Higgsino mass term, contributing to the Higgs mass parameters. For electroweak symmetry breaking, one needs to explain why the $\mu$ term is of order soft mass parameters. This is the so called $\mu$ problem \cite{Kim:1983dt,muproblem2}. R-parity is imposed for baryon and lepton number conservation in MSSM but it does not forbid a large $\mu$ term.
Thus, we need an extended symmetry of R-parity to solve the $\mu$ problem. It has been recently shown that the $Z_4$ R-symmetry provides an elegant solution to the $\mu$ term as the unique symmetry consistent with SO(10) GUT and anomaly-free by a universal Green-Schwarz mechanism \cite{z4r,znr}. On the other hand, the Peccei-Quinn(PQ) symmetry can be also responsible for explaining the smallness of the $\mu$ term \cite{Kim:1983dt,kimnilles}, if it is broken by SUSY breaking only.
In particular, if the PQ symmetry is broken at an intermediate scale, the PQ axion could solve the strong CP problem too \cite{PQ}. 

%inflation with non-minimal coupling

The inflation model using Higgs boson in the Standard Model as the inflaton has recently drawn much attention \cite{Bezrukov:2007ep}.
The key idea is that a quartic potential becomes flat at large field limit due to a non-minimal coupling of the inflaton to the curvature scalar.
The Higgs inflation has been extended to the supersymmetric case in which 
the next-to-MSSM(NMSSM) with a light singlet is necessary as the Higgs inflation occurs along the D-flat direction \cite{jsugra,hmlee,ferrara}. Recently, the supersymmetric inflation with right-handed(RH) sneutrino has been studied in the presence of a non-minimal coupling \cite{newsneutrino}. In this type of inflation models, for the self-coupling of the inflaton candidate to be of order one, a large non-minimal coupling is required to match the COBE normalization of the density perturbation. Thus, there have been an extensive discussion on the unitarity problem due to the large non-minimal coupling \cite{unitarity}. 
During inflation, Higgs inflation looks consistent with the semi-classical approximation, because the unitarity cutoff depends on the background Higgs field value \cite{ferrara,shapo}. However, a UV completion of the Higgs inflation at unitarity scale seems to suggest a change in the form of the Higgs potential with additional interactions \cite{giudicelee}.
Apart from the large non-minimal coupling, the generic feature of Higgs inflation and its variants is that the reheating temperature after inflation is quite high due to a large coupling of the inflaton to the SM \cite{reheating}.
Therefore, there is the gravitino problem in the supersymmetric realizations of Higgs inflation \cite{Khlopov:1984pf}

%sneutrino inflation

In this paper, we consider the singlet extension of the MSSM with right-handed (RH) neutrinos for solving the $\mu$ problem with approximate PQ symmetry.  We assume that supersymmetry breaking is mediated by gauge interaction \cite{Nelson:1993nf,Giudice:1998bp}. The minimal setup for a spontaneous breaking of the PQ symmetry requires the introduction of two SM-singlet flaton fields $X,Y$ with nonzero PQ charges, both of which get intermediate-scale VEVs
{\bf \footnote{Here we assumed that the stabilization of symmetry breaking field is achieved by higher order term(s).}}. 
The flaton $X$ generates a small $\mu$ term by dimension-5 operator while the flaton $Y$ gives large RH sneutrinos masses by renormalizable  couplings. 

The same coupling of the flaton $Y$ to the RH sneutrino provides a flat potential for inflation at large sneutrino field values in the presence of a large non-minimal coupling. It is the quartic coupling that drives sneutrino inflation, in contrast to the early sneutrino inflation models \cite{oldsneutrino} where the sneutrino mass term is responsible for inflation. 
Because of small neutrino Yukawa couplings, the reheating temperature after inflation is much smaller than the one in Higgs inflation. However,  the gravitino problem persists because the bound on the reheating temperature becomes much stronger in gauge mediation.
In our model, thermal inflation is a natural consequence of the flaton $X$,
that couples to extra vector-like states for the radiative symmetry breaking. 
After thermal inflation, the previously produced gravitinos are erased, so is the baryon asymmetry. Moreover, we produce the correct baryon asymmetry via Affleck-Dine(AD) leptogenesis and generate the right amount of dark matter from gravitino. 
Stability of sneutrino inflation requires non-inflaton RH neutrinos of masses to be less than about $\TeV$ scale so they are within the reach of present collider experiments.

The paper is organized as follows:
We first present the model setup to solve the $\mu$ problem of the MSSM. Then,  we discuss the sneutrino inflation in the presence of a non-minimal coupling, addressing the constraints coming from the stability of orthogonal directions to the inflaton.
Next we describe post-inflation cosmology including thermal inflation with the flaton field and baryogenesis and dark matter issues. We also present a concrete UV completion for obtaining the frame function necessary for a stable sneutrino inflation
and comment on the consequence of the PQ symmetry breaking caused by the non-minimal coupling.
Finally the conclusion is drawn. There are four appendices dealing with the stabilization of the flaton and the saxion/axino mass spectra,  
the general framework for Jordan frame supergravity, the computation of the number of efoldings, and the discussion on the critical temperature for AD leptogenesis.

\section{The model}

We consider a similar extension of the MSSM with singlet chiral superfields as in Ref.~\cite{Choi:2011rs}.
In the framework of gauge-mediated supersymmetry breaking \cite{Nelson:1993nf} 
\footnote{
In gravity mediation, the boundary condition for the low energy mass spectrum is given at the Planck scale while the cutoff scale of the theory with largish non-minimal gravitational coupling is much less than Planck scale. So, gravity mediation of supersymmetry breaking is not a proper framework for our study.
}, the model is described by the following superpotential,
\bea \label{W}
W &=& \lambda_u Q H_u \bar{U} + \lambda_d Q H_d \bar{D} + \lambda_e L H_d \bar{E} + \frac{1}{2} \lambda_\mu \frac{X^2 H_u H_d}{\Lambda}
\nonumber \\
&& + \lambda_N L H_u N + \frac{1}{2} \lambda_Y Y N^2
\nonumber \\
&& + \lambda_\Psi X \Psi \bar{\Psi} + \frac{1}{3} \lambda_X \frac{X^3 Y}{\Lambda}
\nonumber \\
&& + \lambda_Z Z \Phi \bar{\Phi}.
\eea
The first line corresponds to the MSSM superpotential where the $\mu$ term is generated by the dimension-5 operator while the second line contains the neutrino Yukawa couplings and RH neutrinos for generating neutrino masses by see-saw mechanism.
The third line is responsible for stabilizing the flatons. When the first term derives the soft mass squared of $X$ to a negative value around the origin by renormalization group running, $X$ is stabilized by the second term.
Here we have introduced extra vector-like states of $SU(5)$, $\Psi$ and $\bar\Psi$, which get soft masses from gauge mediation.  The last line is the messenger sector for gauge mediation, containing another vector-like states of $SU(5)$, $\Phi$ and $\bar\Phi$, and SUSY-breaking field, $Z$ with $\langle Z \rangle = M + \theta^2 F$.
Here we took the cutoff scale to be $\Lambda = M_\planck / \xi_1$ from the sneutrino non-minimal coupling $\xi_1$ (see section \ref{sec:unitarity}) with $M_\planck = 2.4 \times 10^{18} \GeV$ being the reduced Planck mass. The couplings in the first and second line of \eq{W} except $\lambda_\mu$ are understood as $3\times3$ matrices, and $\lambda_Y$ is assumed to be diagonal without loss of generality.

The model possesses the PQ symmetry with charges assigned as in Table \ref{table:U1charges}.
\begin{table}[ht]
\centering
\begin{tabular}{c||c|c|c|c|c|c|c|c|c|c|c|c}
\hline 
& $Q$ & $L$ & $\bar{U}$ & $\bar{D}$ & $\bar{E}$ & $N$ & $X$ & $Y$ & $\Psi$ & $\bar{\Psi}$ & $H_u$ & $H_d$ \\ [0.5ex]
\hline 
PQ & $\frac{3}{2}$ & $-\frac{1}{2}$ & $-\frac{1}{2}$ & $-\frac{1}{2}$ & $\frac{3}{2}$ & $\frac{3}{2}$ & 1 & $-3$ & $-\frac{1}{2}$ & $-\frac{1}{2}$ & $-1$ & $-1$ \\ [0.5ex]
\hline
\end{tabular}
\caption{PQ charges}
\label{table:U1charges}
\end{table}  
This symmetry is actually broken by the non-minimal coupling of RH neutrinos in the frame function (\eq{framefull}).
However, the coupling is relevant only above the cutoff scale $\Lambda$.
Hence we regard the PQ symmetry to be approximate below the cutoff scale and remain a working solution to the $\mu$ problem.
One may attempt to identify the PQ symmetry as the axion solution of strong CP problem with additional $\mathbb{Z}_{24}$ discrete $R$-symmetry \cite{znr}.
However, the non-minimal coupling of the sneutrino with nonzero PQ charge causes too large tadpole contribution to the axion to keep the axion solution (see section \ref{sec:unitarity}), hence it is not plausible to accommodate the axion solution in our minimal setup.

The VEVs of flaton fields are given by   
\bea 
\label{Pvev}
X_0 &\simeq&  3^{1/4} \left( \frac{m_{X}\Lambda}{|\lambda_X|} \right)^{1/2},\\
\label{Qvev1}
Y_0 &\simeq & \frac{1}{3\sqrt{3}}\,\frac{A_{\lambda_X}}{m_X}\,X_0
\eea
where $m_X, A_{\lambda_X}$ are soft mass parameters for the flatons, as given in eqs.~(\ref{softmXsq}) and (\ref{softAX}), respectively. We have determined the mass spectrum in the flaton sector in appendix A.
Then,  the $\mu$ term is generated from the last term in the first line of \eq{W} when the $X$ singlet gets an intermediate-scale VEV,  
\be
\mu = \frac{1}{2} \lambda_\mu \frac{X_0^2}{\Lambda}\simeq \frac{\sqrt{3}}{2}\,\frac{\lambda_\mu}{\lambda_X}\,m_X.
\ee
On the other hand, the large VEV of the $Y$ singlet gives rise to RH sneutrino masses for see-saw mechanism. Integrating out heavy RH neutrinos, one obtains left-handed neutrino mass terms
\be \label{W-nu-mass}
W_{\nu - \mathrm{mass}} = - \frac{1}{2} \frac{\left( LH_u \right)^T \lambda_\nu LH_u}{Y}
\ee
where $\lambda_\nu \equiv \lambda_N \lambda_N ^T \lambda_Y^{-1}$.
Thus, from the see-saw relations for light neutrino masses,
\be \label{LHnu-mass}
m_\nu^{ij} = \frac{1}{2} \lambda_\nu^{ij} \frac{v_u^2}{Y_0},
\ee
we find that the inflaton couplings of Dirac mass term is constrained as
\bea \label{lambdaNI-bound}
\left( \lambda_N \right)_{iI} 
&<& \left( 2 \frac{m_\nu^{ii} Y_0}{v_u^2} \lambda_{YI} \right)^{1/2} \nonumber
\\
&\simeq& 7.9 \times 10^{-6} \left( \frac{m_\nu^{ii}}{10^{-2} \eV} \right)^{1/2} \left( \frac{Y_0}{10^8 \GeV} \right)^{1/2} \left( \frac{\left( \lambda_{YI} \right)}{10^{-3}} \right)^{1/2} 
\eea
where the subscript ``I'' represents inflaton direction. 
On the other hand, as will be shown later in \eq{stability-cond3}, for non-inflaton directions with $j\neq I$,  we find
\be
\label{lambdaN-bound}
\left( \lambda_N \right)_{ij} 
\lesssim 8.0 \times 10^{-7} \left( \frac{m_\nu^{ii}}{10^{-2} \eV} \right)^{1/2} \left( \frac{Y_0}{10^8 \GeV} \right)^{1/2} \left( \frac{\left( \lambda_{Yj} \right)}{10^{-5}} \right)^{1/2}.
\ee
Here we have normalized the neutrino Yukawa couplings, based on the value of $\lambda_{YI}$ from the unitarity at GUT scale and the value of
$\lambda_{Yi\neq I}$ from the stability of non-inflaton sneutrinos, as will be discussed in next section.

\section{Sneutrino inflation} \label{inflation}

In this section, we discuss the chaotic inflation in our model.
To this, we need to specify the K\"ahler potential because the inflation potential depends on the form of the K\"ahler potential at large inflaton values.
Thus, motivated by the Jordan frame supergravity in which the kinetic terms for scalar fields are of canonical form \cite{jsugra,hmlee}, we take the following frame function and the superpotential relevant for sneutrino inflation,
\bea 
\Omega&=&-3+|Y|^2\Big(1-\gamma|Y|^2 - \sum_{i \neq 1} \delta_i |N_i|^2\Big)+\sum_{i=1}^3\Big[|N_i|^2-\frac{3}{2}( \xi_i N_i N_i+{\rm h.c.})\Big], \label{framefull}\\
W&=&\frac{1}{2}\sum_{i=1}^3\lambda_{Yi} Y N_i N_i .\label{superp}
\eea
Here and from now on we use Planck unit.
There are more details on Jordan frame supergravity in appendix B.
Here we have introduced in the frame function, the non-minimal couplings for sneutrinos, $\xi_i$, as well as the higher order terms for the non-inflaton fields, $Y$ and $N_{i\neq 1}$.  The non-minimal coupling becomes dominant at large sneutrino inflaton value, flattening the quartic potential for $N_1$. 
As will be discussed, the higher order terms, $\gamma,\delta_i$, are necessary for the stability of the non-inflaton fields during inflation. A microscopic model for obtaining such higher order terms without spoiling the slow-roll inflation will be discussed in a later section.
We note that the frame function is related to the K\"ahler potential by $\Omega=-3\,e^{-K/3}$.

\subsection{Slow-roll inflation}

Choosing the direction with $Y = N_2 = N_3 = 0$, we obtain the effective action for the sneutrino inflation in Einstein frame \cite{hmlee} as 
\be
\frac{{\cal L}_E}{\sqrt{-g_E}}=\frac{1}{2}R-K_{N_1{\bar N}_1}|\partial_\mu N_1|^2-\frac{\frac{1}{4}|\lambda_{Y1}|^2 |N_1|^4}{(1-\frac{1}{3}|N_1|^2+\frac{1}{2}(\xi_1 N^2_1+{\rm h.c.}))^2}
\ee
where the K\"ahler metric for $N_1$ is
\be
K_{N_1{\bar N}_1}=\frac{1-\frac{1}{2}(\xi_1 N^2_1+{\rm h.c.})+3\xi^2_1|N_1|^2}{(1-\frac{1}{3}|N_1|^2+\frac{1}{2}(\xi_1 N^2_1+{\rm h.c.}))^2}.
\ee
For $\xi_1(3\xi_1-1)|N_1|^2\gg 1$, stabilizing the angular mode of $N_1$, we obtain
the following approximate form of the action,
\be
\frac{{\cal L}_E}{\sqrt{-g_E}}\simeq \frac{1}{2}R-\frac{3\xi_1(\xi_1-\frac{1}{3}) |N_1|^2}{[1+(\xi_1-\frac{1}{3}) |N_1|^2]^2}(\partial_\mu |N_1|)^2-\frac{9|\lambda_{Y1}|  ^2}{4(3\xi_1-1)^2}\Big(1+\frac{3}{(3\xi_1-1) |N_1|^2}\Big)^{-2}.
\ee
Thus, for a canonical scalar field, $\varphi=\frac{1}{a}\ln(1+(\xi_1-\frac{1}{3}) |N_1|^2)$ with $a\equiv \sqrt{\frac{2}{3}-\frac{2}{9\xi_1}}$, the Einstein-frame action becomes
\be
\frac{{\cal L}_E}{\sqrt{-g_E}}\simeq \frac{1}{2}R-\frac{1}{2}(\partial_\mu\varphi)^2-\frac{9|\lambda_{Y1}|^2}{4(3\xi_1-1)^2}\Big(1-e^{-a\varphi}\Big)^2.
\ee
The slow-roll inflation takes place for $e^{-a\varphi}\ll 1$, i.e. 
\be \label{sinf-cond}
(\xi_1-\frac{1}{3}) |N_1|^2\gg 1,
\ee
which implies $\xi \gg 1$ for $|N_1| \lesssim 1$.
Then, the number of efoldings is
\be \label{efoldings}
N_e 
= -\int^e_* \frac{V_E}{\frac{\partial V_E}{\partial\varphi}}d\varphi 
\simeq \frac{1}{2a^2} e^{a\varphi_*}
\ee
where the subscripts $e,*$ mean the end of inflation and the horizon exit.
Moreover, from the slow-roll parameters,
\bea
\epsilon &=& \frac{1}{2}\bigg(\frac{\frac{\partial V_E}{\partial \varphi}}{V_E}\bigg)^2= \frac{2a^2 e^{-2a\varphi}}{(1-e^{-a\varphi})^2}, 
\\
\eta &=& \frac{\frac{\partial^2 V_E}{\partial\varphi^2}}{V_E}= -\frac{2a^2 e^{-a\varphi}(1-2e^{-2a\varphi})}{(1-e^{-a\varphi})^2}.
\eea
we obtain the slow-roll parameters at horizon exit in terms of the number of efoldings as
\be
\epsilon_* \simeq \frac{1}{2 a^2 N_e^2}, 
\quad
\eta_* \simeq - \frac{1}{N_e} 
\ee
From \eq{efoldings}, the field value of inflaton at horizon exit is given by
\be \label{N-exit}
|N_1|(t_*) 
\simeq \sqrt{\frac{2a^2 N_e}{\xi_1}}\, M_\planck
\simeq \left( \frac{N_e}{52} \right)^{1/2} \left( \frac{70}{\xi_1} \right)^{1/2}
\ee
where use is made of $N_e = 52$ as a representative value of efoldings, taking into account of the contribution from thermal inflation (see appendix \ref{sec:efoldings}).
Slow-roll inflation ends when $\epsilon \simeq 1$, hence the field value of inflaton at the end of inflation is given by
\be \label{N-end}
|N_1|(t_e) \simeq \left( \frac{4}{3} \right)^{1/4} \frac{1}{\sqrt{\xi_1}}\,.
\ee

The density perturbation at horizon exit is given by
\be
\Delta^2_{\cal R}
=\frac{V_E}{24\pi^2 \epsilon_*}
\simeq \frac{N_e^2}{8 \pi^2} \frac{|\lambda_{Y 1}|^2}{(3\xi_1-1)^2}.
\ee
Thus, from the COBE normalization, $\delta_H=\frac{2}{5}\Delta_{\cal R}=(1.91\pm 0.17)\cdot 10^{-5}$, we get a constraint on the dimensionless inflation parameters as
\be \label{const-on-lambdaY} 
\lambda_{Y1} \simeq 2.4 \times 10^{-3} \left( \frac{\xi_1}{100} \right).
\ee 
The spectral index and the tensor to scalar ratio are estimated as
\bea
n_s &\equiv& 1 - 6 \epsilon + 2 \eta \simeq 0.96,
\\
r &=& 16 \epsilon \simeq 4.4 \times 10^{-3}. 
\eea
The results are consistent with the observed values by WMAP \cite{Komatsu:2010fb}.
We note that the spectral index is smaller than the one in Higgs inflation due to the thermal inflation and the tensor to scalar ratio remains small.

\subsection{Stability of the non-inflaton fields}

During inflation (i.e., $|\xi_1 N_1^2| \gg 1$), along the direction with $N_2=N_3=0$, the Einstein-frame potential becomes \cite{hmlee}
\be
V_E \simeq  \frac{1}{4} \frac{\lambda^2_{Y1}}{\xi^2_1} \left[ 1 + \left( 4 \gamma - \frac{2}{3\xi_1 |N_1|^2} \right) |Y|^2 \right] \label{stableY}
\ee
while along the direction with $Y=0$ the potential becomes 
\be
V_E \simeq  \frac{1}{4} \frac{\lambda^2_{Y1}}{\xi^2_1} \left[ 1 +\sum_{i\neq 1} \delta_i |N_i|^2- \sum_{i \neq 1} \left| \frac{\lambda_{Yi}}{\lambda_{Y1} N_1^2} \right| (N^2_i+{\bar N}^2_i) \right].
\ee
Therefore, requiring that non-inflaton directions are stable at least until the end of inflation, we find constraints,
\bea
\label{stability-cond1}
\gamma &>& \frac{1}{6\xi_1|N_1|^2(t_e)} \simeq 0.1,
\\
\label{stability-cond2}
\delta_i &>& 2\left| \frac{\lambda_{Yi}}{\lambda_{Y1} N_1^2(t_e)} \right| \simeq 7\times 10^4 |\lambda_{Yi}|
\eea
where use is made of eqs.~(\ref{N-end}) and (\ref{const-on-lambdaY}). It is theoretically natural to expect that $\gamma$, $\delta_i \lesssim 1$ unless there is any special mechanism to generate those terms at a scale much lower than the Planck scale.
Hence \eq{stability-cond2} becomes  or non-inflaton directions
\footnote{
For $\lambda_{Yi} \sim 10^{-5}$, soon after the end of inflation, inflaton would be destabilized along the direction of non-inflaton direction(s).
As a result, order unity fractional energy density of inflaton might be transmitted to those directions.
But, it does not cause any problem as long as RH-(s)neutrinos decay before the time of Big Bang nucleosynthesis.
}
\be \label{stability-cond3}
\lambda_{Yi} \lesssim 10^{-5}.
\ee 
Therefore, for the Y flaton VEV of order $10^8\,{\rm GeV}$, the non-inflaton sneutrinos or neutrinos must be less than $100\,{\rm GeV}$.

\section{Post inflation}
\label{sec:postinflation}

After inflation, we confront a nontrivial and involved dynamics of the inflaton and the flatons, determining the post-inflation evolution of the universe.
In this section, we discuss post-inflation cosmology, including thermal inflation, baryogenesis and dark matter issues.

\subsection{Thermal inflation}

The thermal history in our model after inflation is rather complicated.
To help readers have a clearer picture, we list the temperatures at various epochs critical in our argument in the order of time. 
\begin{itemize}
\item $T_\mathrm{b}$: Thermal inflation begins.
\item $T_\mathrm{R}$: Inflaton decay is completed.
\item $T_{LH_u}$: $LH_u$ flat direction is destabilized from the origin.
\item $T_\mathrm{c}$: Thermal inflation ends as $X$ is destabilized from the origin.
\item $T_\mathrm{d}$: Flaton ($X$) decay is completed.
\end{itemize}

After inflation, the inflaton oscillates coherently with initial amplitude larger than GUT scale, causing the preheating \cite{preheating} of particles coupled to it.
Without getting into the complicated details of the preheating process, we simply estimate the reheating temperature from the perturbative decay, which will be enough for subsequent discussions.
The perturbative decay of the sneutrino inflaton occurs due to the neutrino Yukawa couplings.
When the inflaton oscillates in the quartic potential, the effective inflaton mass is given by $m_I = \sqrt{3/2} \lambda_{YI} N_I$.
Thus, the inflaton decay rate due to the neutrino Yukawa couplings is 
\be
\Gamma_I = \frac{\sqrt{3/2}}{8 \pi} \sum_i |\left( \lambda_N \right)_{iI}|^2 \lambda_{YI} N_I.
\ee
Equating the decay rate to the expansion rate of the universe, we find that the reheating temperature is bounded as
\bea \label{TR}
T_\mathrm{R} 
&\gtrsim& \left( \frac{\pi^2}{30} g_*(T_\mathrm{R}) \right)^{-1/4} \frac{3}{8 \pi} \sum_i \left( \lambda_N \right)_{iI}^2 \left( \lambda_{YI} \right)^{1/2}
\nonumber \\
&\simeq& 3 \times 10^5 \GeV \left( \frac{g_*(T_\mathrm{R})}{200} \right)^{-1/4} \left( \frac{\left( \lambda_N \right)_{iI}}{10^{-5}} \right)^2 \left( \frac{\left( \lambda_{YI} \right)}{10^{-3}} \right)^{1/2}\,.
\eea 
Therefore, the gravitino problem \cite{Khlopov:1984pf} is present unless the gravitino mass is larger than about a few $\MeV$ \cite{Moroi:1993mb}.
On the other hand, as will be described subsequently, thermal inflation \cite{Lyth:1995hj} is a natural consequence of our model so that gravitino problem disappears for the whole range of the gravitino mass possible in gauge mediation.

Thermal inflation begins when the energy density of radiation becomes comparable to $V_0$ while $X$ is still held around the origin due to thermal effect
\footnote{The flaton $X$ can get a large positive Hubble scale mass-squared originated from gravity mediation which will hold $X$ around the origin during inflation.
After inflation, preheating and partial decay of inflaton raise up the temperature of the universe above the symmetry breaking scale of PQ-symmetry, hence $X$ can be still in the symmetric phase around the origin.
Therefore, in our scenario, thermal inflation is inevitable.
On the other hand, a higher order correction for $X$ in the K\"ahler potential may lead to a tachyonic mass for $X$ at the origin. Then, thermal inflation might not occur, depending on the maximal temperature after inflation.
}. As seen from the flaton potential \eq{Vphi}, along the $X=0$ direction, the flaton $Y$ is also stable at the origin, keeping trapped at the origin by the inflaton-induced mass term in \eq{stableY} during inflation and by gravity-mediation effect after inflation. As the $X$ flaton gets destabilized, the $Y$ flaton also rolls out to the true minimum due to the interaction with $X$ flaton.
Here the vacuum energy $V_0$ is estimated from requiring a zero cosmological constant at the vacuum as
\be \label{V0}
V_0 \simeq \frac{2}{3} m_{X}^2 X_0^2 \simeq \frac{2 \sqrt{3}}{3} \frac{m_{X}^3 \Lambda}{|\lambda_X|}.
\ee
The temperature at the beginning of thermal inflation is 
\be
T_\mathrm{b} \sim V_0^{1/4} \GeV \sim 10^{6.5} \GeV \left( \frac{m_X(X_0)}{1 \TeV} \right)^{1/2} \left( \frac{X_0}{10^{10} \GeV} \right)^{1/2}\,.
\ee
This is higher than $T_\mathrm{R}$, meaning that thermal inflation begins before inflaton decay is completed.
Therefore, $T_\mathrm{b}$ is the temperature not of standard model particles, but of inflaton which behaves like radiation after inflation.

Thermal inflation ends as $X$ is destabilized from the origin.
If the supersymmetric masses of RH-(s)neutrinos are negligible (i.e., $m_{3/2} \ll m_\mathrm{soft} / \xi$),
%\footnote{That is, $m_{3/2} \ll m_\mathrm{soft} / \xi$.}
, the critical temperature of the destabilization is given by
\be \label{Tc}
T_\mathrm{c}\simeq  \frac{m_X(0)}{\beta_X}
\ee
where $m_X(0)$ is given by \eq{softmXsq} and $\beta^2_X=\frac{1}{4}N_{\Psi}\sum|\lambda_{\Psi_i}|^2$.
Therefore, the total number of $e$-foldings of thermal inflation is 
\be
N_\mathrm{TI} 
= \ln \frac{a_\mathrm{c}}{a_\mathrm{b}}
\simeq 7.0 + \ln \left( \frac{T_\mathrm{b}}{10^6 \GeV} \right) - \ln \left( \frac{T_\mathrm{c}}{1 \TeV} \right)\,.
\ee 
 
Soon after thermal inflation, the coherent oscillation of $X$ becomes dominant, and its eventual decay reheats the Universe, releasing huge amount of entropy.
For $m_x > 2 m_h$ with $m_x$ being the physical flaton mass and $m_h$ being the light Higgs boson, the decay rate of $X$ is 
\be
\Gamma_{X \to \mathrm{SM}}
\simeq 
\frac{1}{4 \pi} \left( 1 - \frac{|B|^2}{m_A^2} \right)^2 \left( \frac{|\mu|^4}{m_x X_0^2} \right) \left( 1 - \frac{4 m_h^2}{m_x^2} \right)^{1/2}
\ee
where $B, m_A$ are the B-term for Higgs doublets and the CP-odd Higgs mass, respectively.
Then, the decay temperature of the flaton $X$ is 
\bea \label{Tdphi}
T_\mathrm{d}
&\equiv& \left( \frac{\pi^2}{15} g_*(T_{\mathrm{d} X}) \right)^{-1/4} \left( \Gamma_{X \to \mathrm{SM}} \Gamma_X \right)^{1/4} M_\planck^{1/2} \nonumber
\\
&\simeq& 408 \GeV \left( \frac{\mu}{1 \TeV} \right)^2 \left( \frac{1 \TeV}{m_x} \right)^{1/2} \left( \frac{10^{10} \GeV}{X_0} \right)
\eea
where we have used $g_*(T_\mathrm{d}) = 200$, $B=200 \GeV$, $m_A = 1 \TeV$ and $m_h = 120 \GeV$ in the second line
\footnote{A large enough $B$ term can be obtained by large renormalization group running if messenger mass is of intermediate scale \cite{Choi:2011rs}.}.
The entropy released in the decay of $|X|$ leads to a dilution factor, 
\bea 
\Delta_X \label{phidfactor}
&=& \frac{V_0}{T_\mathrm{c}^3 T_\mathrm{d}} \nonumber
\\
&\simeq& 2 \times 10^{13} \left( \frac{m_x}{T_\mathrm{c}} \right)^2 \left( \frac{1 \TeV}{T_\mathrm{c}} \right) \left( \frac{1 \TeV}{T_\mathrm{d}} \right) \left( \frac{X_0}{10^{10} \GeV} \right)^2
\eea
where we have ignored the fractional energy loss of flaton to no-SM particles since it does not make any change in our argument.
Note that the dilution is large enough to remove gravitino problem caused by high reheating temperature after primordial inflation.

Our model has two other oscillating scalar fields which are mostly ${\rm Re}(Y)$ and ${\rm Im}(Y)$.
Although they have a mass comparable to $m_x$, their energy densities are suppressed by $Y_0^2 / X_0^2 = {\cal O}(g_s^4\lambda_X^2/(8\pi^3)^2)$ compared to that of $|X|$, and are not dominant when they decay. 
Therefore they do not give a significant impact on the cosmological evolution after thermal inflation.

\subsection{Baryogenesis}

In the presence of thermal inflation, pre-existing baryon/lepton asymmetry can not endure the large dilution caused by the entropy release of thermal inflation \footnote{It is possible to have baryogenesis before thermal inflation, provided that Affleck-Dine field generates a sufficiently large initial $n_B/s$ and it decays after thermal inflation \cite{gouvea}. }.
Hence we have to regenerate baryon/lepton asymmetry after thermal inflation
\cite{Stewart:1996ai, Jeong:2004hy,Felder:2007iz, Kim:2008yu, Choi:2011rs}.

The condition for a late-time Affleck-Dine leptogenesis is $T_c < T_{LH_u}$, under which the AD field is destabilized earlier than the flaton $X$.
As shown in appendix D, this condition is fulfilled in our model so the AD leptogenesis works in the same way as in the model of Ref.~\cite{Choi:2011rs}.
We restrict ourselves to the flatons, $L_iH_u$ and $H_u H_d$ flat directions, parametrized by
\be
L_i=(0,l_i)^T,\quad H_u=(h_u,0)^T,\quad H_d=(0,h_d)^T.
\ee
At large flaton field values $|X|, |Y|\gg m_{\rm soft}$, we can integrate out the RH neutrinos to get the effective potential as follows,
\bea
V&=&m^2_{L} |l|^2+m^2_{H_u}|h_u|^2+m^2_{H_d}|h_d|^2-m^2_X |X|^2+m^2_Y |Y|^2 \nonumber \\
&&+\frac{1}{2} A_\mu\lambda_\mu\frac{X^2h_u h_d}{\Lambda}-\frac{1}{2}A_N\frac{\lambda^2_N}{\lambda_Y Y}\,(l \,h_u)^2+\frac{1}{3}A_{\lambda_X}\lambda_X\frac{X^3Y}{\Lambda}+{\rm c.c.} \nonumber \\
&&+\bigg|\frac{1}{2}\lambda_\mu\frac{X^2 h_d}{\Lambda}-\frac{\lambda^2_N}{\lambda_Y Y}\,l\,(l\, h_u)\bigg|^2+\bigg|\frac{1}{2}\lambda_\mu \frac{X^2 h_u}{\Lambda} \bigg|^2 \nonumber \\
&&+\bigg|\lambda_\mu \frac{Xh_u h_d}{\Lambda}+\lambda_X\frac{X^2 Y}{\Lambda}\bigg|^2 +\bigg|\frac{1}{2}\frac{\lambda^2_N}{\lambda_Y Y^2}(l\, h_u)^2+\frac{1}{3}\lambda_X\frac{X^3}{\Lambda} \bigg|^2. \label{ADpot}
\eea  
The $L_iH_u$ flat direction rolls out to non-zero value at a temperature $T \sim m_{L_iH_u}$
\footnote{
The condensation of $LH_u$ and $H_uH_d$ dumps some amount of energy before the end of thermal inflation.
As a result, the background temperature is raised up, extending thermal inflation a couple of efoldings more \cite{Kim:2008yu}.
}.
It is stabilized by the radiative effect rather than the small tree-level higher order term, hence the stabilized value depends on the messenger scale. 
%{\bf Due to the interplay between the soft mass terms and the higher order term in the potential (\ref{ADpot}), the $L_iH_u$ flat direction is stabilized at
%\be
%|l|^2\sim |h_u|^2\sim  \frac{|\lambda_Y Y| |m^2_{LH_u}|}{|\lambda^2_N|}
%\ee
%where $m^2_{LH_u}$ is given in \eq{lhumass}.}
From a numerical calculation, we found that $L_iH_u$ is stabilized at $|\ell_0| \sim \mathcal{O}(10^{6-7}) \GeV$, for $m_{3/2} \sim 100 \keV$, which is of our interest with respect to dark matter.
%It is the expected initial value of $LH_u$ before it rolls into the origin.
When $X$ and $Y$ flatons eventually reach the true vacuum values,
the $\mu$ term is generated, providing additional masses to $L_iH_u$ and $H_uH_d$ flat directions.
As a result, those flat directions are brought back into the origin.
In this process, the $X$-dependent $CP$-violating term of $L_iH_u$ causes an angular kick for the motion of $L_iH_u$ so that Affleck-Dine leptogenesis works.

To be conservative, however, one has to pay attention to the fact that in gauge mediation, $H_uH_d$ is likely to be destabilized earlier than $L_iH_u$ while the $\mu$ term is absent.
This implies that $L_iH_u$ flat directions could obtain a large mass due to the neutrino Yukawa coupling, $\lambda_N$.
Hence, in order to make late-time Affleck-Dine leptogenesis work, all the entries of $\lambda_N$ associated with a certain flavor of lepton douplets (say $L_i$) should satisfy a condition 
\be \label{LHu-flat-cond1}
\left| \lambda_N^{ij} \right| \ll \frac{m_{L_iH_u}}{\langle H_u \rangle} \sim 5 \times 10^{-4} \left( \frac{m_{L_iH_u}}{500 \GeV} \right) \left( \frac{10^6 \GeV}{\langle H_u \rangle} \right)
\ee
so that the mass contribution to the flavor $L_i$ due to the early destabilization of $H_u H_d$ is small enough not to hold $L_i H_u$ around the origin.
Note that the above condition is automatically satisfied for \eq{lambdaN-bound} with \eq{stability-cond3}.

The generated lepton number asymmetry is expected to be conserved by the help of rapid preheating of $X$ and $L_iH_u$ flat directions \cite{Felder:2007iz,Kim:2008yu}, and finally converted to baryon asymmetry through the sphaleron process \cite{Kuzmin:1985mm}.
Including the dilution due to entropy release in the eventual decay of $|X|$, the resulting baryon asymmetry at present is estimated as \cite{Jeong:2004hy}
\be \label{YBpresent}
\frac{n_B}{s} \sim \frac{n_B}{n_x} \frac{T_\mathrm{d}}{m_x} \sim \frac{n_L}{n_\mathrm{AD}} \frac{n_\mathrm{AD}}{n_x} \frac{T_\mathrm{d}}{m_x} \sim \frac{n_L}{n_\mathrm{AD}} \frac{m_{L_iH_u}}{m_x} \left( \frac{|l_0|}{X_0} \right)^2 \frac{T_\mathrm{d}}{m_x}
\ee
where $n_x$, $n_L$ and $n_\mathrm{AD}$ are number densities of $|X|$, lepton asymmetry and AD field, respectively.
For a small $CP$-violating phase, $\delta\ll 1$, the conserved lepton asymmetry can be expressed as 
\be
n_L \sim \alpha \, \delta \, m_\theta |\ell_0|^2
\ee
where $\alpha \sim 0.1$ is the efficiency factor of conserving the generated asymmetry \cite{Felder:2007iz,Kim:2008yu}, and $m_\theta$ is the mass of the angular mode of the $LH_u$ direction when it is lifted up and starts to roll in.
We find
%\be
%m_\theta^2 \sim \mu\, \frac{\lambda^2_N\lambda_X}{\lambda_Y}\frac{X_0}{Y^2_0}\, |l_0|^2.
%\ee
\be
m_\theta^2 \sim \mu\, \left( \frac{\lambda_X X_0}{\lambda_\mu Y_0} \right) \frac{\lambda^2_N}{\lambda_Y Y_0}\, |l_0|^2.
\ee
Hence   
\be
\frac{n_L}{n_\mathrm{AD}} 
\sim \alpha \, \delta \, \left( \frac{m_\theta}{m_{L_iH_u}} \right) 
= 10^{-3} \left( \frac{\alpha}{0.1} \right) \left( \frac{\delta}{0.1} \right) \left( \frac{m_\theta}{50 \GeV} \right) \left( \frac{500 \GeV}{m_{L_iH_u}} \right). 
\ee
and 
\be \label{nBs}
\frac{n_B}{s} \sim 10^{-9} \left( \frac{n_L / n_\mathrm{AD}}{10^{-3}} \right) \left( \frac{m_{LH_u}}{m_x} \right) \left( \frac{|\ell_0| / X_0}{10^{-3}} \right)^2 \left( \frac{T_\mathrm{d}}{1 \TeV} \right) \left( \frac{1 \TeV}{m_x} \right). 
\ee
Therefore, the obtained baryon asymmetry can be consistent with the observation within the uncertainties of involved parameters,

\subsection{Dark matter}

In our model, the gravitino is the lightest supersymmetric particle as it is typical in gauge-mediation, hence it is a good candidate of dark matter at present.
For the decay temperature $T_\mathrm{d} \sim \mathcal{O}(1) \TeV$ after inflation, the gravitinos can be produced from the thermal scattering and decay of MSSM particles and provide a right amount of present dark matter abundanc, provided that \cite{Moroi:1993mb}
\be 
m_{3/2} \sim \mathcal{O}(100) \keV.
\ee
Gravitinos can be also produced non-thermally from the decay of flatons and heavy flatinos.
In this case, gravitinos are expected to be warm unless the masses of flaton and flatino are larger than about $1 \TeV$.
However, if flatinos decay to the ordinary lightest supersymmetric particle(OLSP), the non-thermal production of gravitinos can be negligible \cite{Choi:2011rs, Chun:2011zd}. Therefore, the flatino mass is constrained as
\be \label{flatino-massbound}
m_{f_{1,2}} > m_h + m_{\tilde{B}}.
\ee
Based on \eqs{softmXsq0} {flatinomass}, \eq{flatino-massbound} can be satisfied for $\lambda_\Psi \sim 1$.

\section{A UV completion of the frame function}

In this section, we propose a simple UV completion of the frame function with higher order terms that we considered in the previous sections.
It has been shown that a successful chaotic inflation is possible in Jordan frame supergravity, because integating out heavy fields leads to a necessary higher order term in the one-loop frame function for the stability of the non-inflaton field  \cite{hmlee}.

Following the similar line of the discussion in Ref.~\cite{hmlee}, we introduce four heavy chiral superfields, $\Phi_a (a=1,2,3,4)$ with the following couplings to the non-inflaton sector up to dimension-5 operator,
\be
W=\frac{1}{2}\kappa Y\Phi^2_1+M_1\Phi_1 \Phi_2 +\frac{1}{2}\alpha_i N_{i\neq 1}\Phi^2_3+M_2 \Phi_3 \Phi_4+\frac{\rho_i}{2\Lambda}\, Y N_{i\neq 1}\Phi^2_4.
\ee
In this UV completion, we assume that the frame function for the inflation sector is of the minimal form as follows,
\be
\Omega=-3+|Y|^2+\sum_{i=1}^3\Big[|N_i|^2-\frac{3}{2}( \xi_i N_i N_i+{\rm h.c.})\Big]+\sum_{a=1}^4|\Phi_a|^2.
\ee
The PQ charges and $Z_2$-parities are assigned in Table \ref{table:UVcharges}.
\begin{table}[ht]
\centering
\begin{tabular}{|c||c|c|c|c|c|c|c|}
\hline 
& $Y$ & $N_1$ & $N_{i\neq 1}$ & $\Phi_1$ & $\Phi_2$ & $\Phi_3$ & $\Phi_4$  \\ [0.5ex]
\hline 
PQ & $-3$ & $\frac{3}{2}$ & $\frac{3}{2}$ & $\frac{3}{2}$ & $-\frac{3}{2}$ & $-\frac{3}{4}$ & $\frac{3}{4}$  \\ [0.5ex] \hline 
$Z_2$ & $+1$ &$+1$ & $+1$ & $-1$ & $-1$ & $-1$ & $-1$ \\ [0.5ex] 
\hline
\end{tabular}
\caption{PQ charges and $Z_2$ parities in a UV completion.}
\label{table:UVcharges}
\end{table}  
Here we note that PQ symmetry and $Z_2$-parity only does not distinguish between $N_1$ and $N_{i\neq 1}$ so there would appear similar couplings of the inflaton sneutrino to the heavy fields, $\Phi_3$ and $\Phi_4$, as the ones for non-inflaton sneutrinos. Then, the inflaton would be sensitive to those couplings to the heavy fields. However, suppose that in extra dimensions, heavy fields and non-inflaton sneutrinos are localized on the hidden brane while inflaton sneutrino and the rest fields of our model are localized on the visible brane. In this case, the direct couplings between the inflaton sneutrino and the heavy fields are geometrically suppressed. Moreover, the small masses of RH neutrinos corresponding to the non-inflaton sneutrinos can be understood as well. 

Since the scalar fields are conformally coupled to the curvature scalar in Jordan frame supergravity \cite{hmlee}, only fermions contribute to the one-loop frame function.
Assuming that the heavy fields do not have VEVs and integrating out the heavy fields, we obtain the renormalized one-loop frame function in terms of the fermion mass eigenvalues as follows,
\bea
\Delta\Omega&=&-\frac{1}{32\pi^2}\sum_{a=1}^4 m^2_{F,a}\ln\Big(\frac{m^2_{F,a}}{\mu^2}\Big) \nonumber \\
&\simeq& -\frac{1}{32\pi^2}\bigg\{2M^2_1 \ln\Big(\frac{M^2_1}{\mu^2}\Big)+\Big[\ln\Big(\frac{M^2_1}{\mu^2}\Big)+2\Big]\kappa^2 |Y|^2+\frac{\kappa^4|Y|^4}{6M^2_1} \nonumber \\
&&+2M^2_2 \ln\Big(\frac{M^2_2}{\mu^2}\Big)+\Big[\ln\Big(\frac{M^2_2}{\mu^2}\Big)+2\Big]\Big(\alpha^2_i |N_i|^2+\frac{\rho^2_i|YN_i|^2}{\Lambda^2}\Big) \nonumber \\
&&+\frac{\alpha^4_i|N_i|^4}{6M^2_2}+\frac{\kappa\rho_i}{\Lambda}(YN^2_i+Y^\dagger N^{\dagger2}_i)+{\cal O}\bigg(\frac{|Y|^2 |N|^4_i}{M^2_2\Lambda^2}\bigg)\bigg\}.  \label{oneloopframe}
\eea
Therefore, as compared to eq.~(\ref{framefull}), we have derived the desired higher order terms for the stable $Y$ and non-inflaton sneutrinos $N_{i\neq 1}$ as
\be
\gamma=\frac{\kappa^4}{192\pi^2M^2_1}, \quad \delta_i=\frac{\rho^2_i}{32\pi^2\Lambda^2}\Big[\ln\Big(\frac{M^2_2}{\mu^2}\Big)+2\Big].
\ee
We note that the fact that the $\delta_i$ parameters depend on the renormalization scale $\mu$ indicates that a new counter term $|Y N_i|^2$ in the frame function is necessary as a consequence of the non-renormalizable coupling $\rho_i$ in the superpotential.
In addition to the above terms, there is a renormalization of the Planck mass by $M^2_i\ln(M^2_i/\mu^2)$ terms;
there are quadratic terms for $Y$ and $N_i$, leading to the wave function renormalizations; the quartic terms for non-inflaton sneutrinos are harmless for inflation. Finally, the (anti-)holomorphic term in the last line of eq.~(\ref{oneloopframe}) does not modify either the kinetic terms or the potential in Jordan frame and it does not affect the stability of non-inflaton fields. However, if there exists a nonzero coupling $\alpha_1$ for the inflaton sneutrino such as $\alpha_i$, the loop-induced quartic term, $|N_1|^4$, in the frame function, would be safe only if it is suppressed as compared to the non-minimal coupling, that is, $|N_1|\ll \frac{1}{\alpha_1}\sqrt{576\xi_1\pi^2}\ M_1$. If the heavy field mass is $M_1\sim\Lambda=\frac{1}{\xi_1}$, for $\xi_1\sim 100$ and $\alpha_1\sim 1$,  the bound on the inflaton field value would be $|N_1|\ll 7$, which is close to the inflation field value at horizon exit in eq.~(\ref{N-exit}).

\section{Non-minimal coupling and PQ symmetry breaking}
\label{sec:unitarity}

The non-minimal coupling to gravity induces a new effective interaction between the graviton and the scalar field, which gives rise to the unitarity bound on the maximum energy scale. In our case, the non-minimal coupling, $F=\xi_1 N^2_1$, gives rise to the effective interaction term in the Jordan frame,
\be
{\cal L}_{\rm eff}\simeq \Big(\xi_1 N^2_1 + {\rm h.c.}\Big)\Box h^\mu_\mu \label{effint}
\ee
where $h^\mu_\mu$ is the trace part of the graviton.
Thus, the upper bound allowed by unitarity on the new-physics scale \cite{unitarity} is given by $\Lambda\simeq \frac{1}{\xi_1}$.
However, it has been shown \cite{shapo} that during inflation, the unitarity scale is as high as $1/\sqrt{\xi_1}$, which is higher than the one in the vacuum, $\Lambda$, for a large $\xi_1$. Nonetheless, in a UV complete model of the Higgs inflation \cite{giudicelee}, new physics entering at unitarity scale in the vacuum has been shown to interfere the inflation with a large non-minimal coupling such that the inflation energy depends on the unknown coupling of new physics. 

The Hubble scale during inflation is approximately given by $H\simeq \frac{|\lambda_{Y 1}|}{6\xi_1}$. 
Taking $\Lambda$ to be the maximum energy scale, we must have $H\ll \Lambda$, resulting in $|\lambda_{Y 1}|\ll \frac{3}{2}$. This is consistent with the fact that with a small self-coupling of the inflaton, the inflation energy is less sensitive to the unknown coupling at unitarity scale \cite{giudicelee}. 
Suppose that $|\lambda_{Y1}|=0.01$. Then, from eq.~(\ref{const-on-lambdaY}), we need to take the non-minimal coupling to be 
$\xi\simeq 42$. In this case, the quantum gravity scale becomes $\Lambda\simeq 0.01 \sim 10^{16}$ GeV, which is close to the GUT scale such that we can trust the perturbative unification of gauge couplings.

On the other hand, the non-minimal coupling $\xi_1$ breaks the PQ symmetry explicitly.
Thus, in the effective theory below the unitarity scale, the PQ symmetry should appear as an accidental symmetry. In gravity mediation, the non-minimal coupling generates an effective supersymmetric mass for the RH neutrino chiral superfield containing the inflaton,
\be
W_\nu=\frac{3}{2} m_{3/2}\xi_1 N_1 N_1.
\ee
In the presence of the above effective supersymmetric mass term, the B-term for the RH sneutrino is also generated as $V_{B_\nu}=\frac{3}{2}B_\nu m_{3/2}\xi_1 N_1N_1$. Then, combining the trilinear soft mass, $A_Y \lambda_{Y 1}Y  N_1 N_1$, with the B-term for $N_1$, one would get the one-loop tadpole term for the flaton $Y$: 
\be \label{tadpole}
\Delta V(Y)\sim \frac{\lambda_{Y1}}{16\pi^2} A_Y B_\nu m_{3/2}\xi_1 \log(\Lambda^2/M_1^2)\,Y
\ee
where $\Lambda$ is the unitarity cutoff and $M_1=\lambda_{Y1} \langle Y\rangle$. 
For $A_Y \sim B_\nu\sim m_{3/2}$ in gravity mediation, the tadpole term would be unacceptably too large for the DFSZ axion solution \cite{dfsz} to strong CP problem to be valid. For the axion potential to be minimized at ${\bar\theta}<10^{-9}$, the gravitino mass is constrained as $m_{3/2} < (\frac{10^2}{\xi_1})^{2/3} \, 100 \eV$ for $X_0 \sim 10^{10} \GeV$ with $Y_0 / X_0 \sim 10^{-2}$.

In gauge mediation, 
the PQ symmetry breaking is realized by the tachyonic mass of the flaton induced by the coupling to extra vector-like states, $\lambda_\Psi$.
Meanwhile, choosing $\xi \sim 100$ at the lowest possible value, the axion solution demands $m_{3/2} \lesssim 100 \eV$, which corresponds to the messenger scale, $M \lesssim 10^{6-7} \GeV$. 
This implies that the coupling $\lambda_\Psi$ should be less than about $\mathcal{O}(10^{-3})$ in order for extra vector-like states to contribute to the scalar soft mass of the flaton.
Such a small coupling leads to the flaton of $\GeV$ or sub-$\GeV$ scale mass and results in the flaton decay temperature of similar scale.
The only plausible scenario for baryogenesis in this case might be the late-time leptogenesis after thermal inflation we have considered here
\footnote{
Since $T_\mathrm{d} \sim \mathcal{O}(1) \TeV$, electroweak baryogenesis \cite{Trodden:1998ym} might be considered.
However, our model is practically the MSSM at low energy, hence electroweak baryogenesis would not be able to generate a right amount of baryon asymmetry \cite{Carena:2002ss}.
}.
However, the resulting baryon asymmetry is expected to be too small due to a quite small angular curvature of the potential for the Affleck-Dine field.  
Moreover, it is difficult to obtain enough amount of dark matter if it consists of gravitinos and axions.
Therefore, even in gauge-mediation, the axion solution with PQ symmetry would be incompatible with post-inflation cosmology in the presence of the non-minimal coupling.  To the axion solution, we need to rely on a type of KSVZ axion models \cite{ksvz} in which the MSSM fields including Higgs doublets and RH neutrinos are neutral under the new global symmetry such that the non-minimal couplings for sneutrinos respect the new global symmetry. As discussed in the previous section, the gravitino mass must be of order $100$ keV for a correct dark matter abundance but the one-loop tadpole term (\ref{tadpole}) affects little the mass spectrum in the flaton sector given in appendix A, apart from the PQ axion.

On the other hand, the PQ symmetry remains the solution to the $\mu$ problem even with the tadpole term, because the PQ breaking VEVs are not changed much if $m_{3/2} \ll \left[ 16\pi^2 m^2_{\rm soft} (X_0/Y_0) f_{PQ}/\xi_1 \right]^{1/3}\sim (\frac{10^4}{\xi_1})^{1/3} 50 \,{\rm TeV}$ for $f_{PQ}\sim 10^{10}$ GeV and $m_{\rm soft}\sim 100$ GeV. That is, the gravitino of 100 GeV or even higher mass is consistent with the $\mu$ term of order $m_{\rm soft}$.
Therefore, even in gravity mediation, the PQ breakdown with the non-minimal coupling would be safe for solving the $\mu$ problem.

\section{Conclusion}

We have considered the sneutrino inflation and post-inflation cosmology in Jordan frame supergravity, based on the singlet extension of the MSSM.
The model is characterized by the superpotential (\eq{W}) and the frame function (\eq{framefull}) in gauge mediated supersymmetry breaking. 
It provides heavy right-handed neutrino masses and the $\mu$ term by the vacuum expectation values of singlets, the flatons.
We have realized a stable sneutrino inflation by means of a non-minimal gravity coupling in the frame function. Higher order terms in the frame function ensure the stability of non-inflaton fields. We also proposed a simple UV completion in which the necessary higher order terms in the frame function are generated by the couplings to heavy fields. But, we found that a distinction between the inflaton sneutrino and the non-inflaton sneutrions is necessary in order not to generate a dangerous higher order term for the inflaton. We argued that a geometric separation between the inflaton sneutrino and the non-inflaton sneutrions in extra dimensions can ensure the stability of non-inflaton fields through their couplings to heavy fields while keeping the slow-roll inflation. 

The reheating temperature after inflation is expected to be larger than $\mathcal{O}(10^5) \GeV$ so gravitinos could be overproduced depending on the mass of the gravitino in gauge mediation.
But, the existence of the flat direction for PQ symmetry breaking gives rise to thermal inflation so that the gravitino problem is solved. 
Thermal inflation ends by symmetry breaking phase transition, triggering Affleck-Dine leptogenesis by generating the $\mu$ term, and resulting in baryon asymmetry within the right range to match the present observation.
The reheating temperature after thermal inflation is of $\mathcal{O}(1) \TeV$, so the gravitino provides the right amount of dark matter if it has mass of $\mathcal{O}(100) \keV$.
Contrary to most of the known inflation scenarios, the successful inflation and post-inflation cosmology tightly constrains the model parameters so that non-inflaton sneutrino directions are constrained to have supersymmetric masses less than $\mathcal{O}(1) \TeV$.
Importantly, a natural realization of late-time Affleck-Dine leptogenesis after thermal inflation has been made without any further assumption.
The spectral index predicted in our scenario is $n_s \simeq 0.96$ due the additional efoldings from thermal inflation. 
This is a clear difference from the original Higgs inflation and its variants where thermal inflation is absent.

\acknowledgments
YGK is supported by the Basic Science Research Program through the National Research Foundation of Korea
(NRF) funded by the Korean Ministry of Education, Science and Technology (2010-0010312).
HML is supported by CERN-Korean fellowship.

\appendix

\section{Flaton potential and mass spectrum} 
\label{sec:mass}

The potential for the flatons, $X$ and $Y$, is given by \footnote{The soft mass squared of $Y$ is dominantly from gravity-mediation effect, i.e., $m^2_Y \sim m^2_{3/2}$.
It is positive even under RG-running, since radiative correction is negligible due to smallness of Yukawa coupling.}
\be \label{Vphi}
V(X) = V_0 - m_X^2 |X|^2 + m_Y^2 |Y|^2+\left( \frac{1}{3} A_{\lambda_X} \lambda_X \frac{X^3 Y}{\Lambda} + \mathrm{c.c.} \right) + \left|\frac{\lambda_X X^3}{3 \Lambda}\right|^2 + \left|\frac{\lambda_X X^2 Y}{\Lambda}\right|^2
\ee
where $-m_X^2$ and $m_Y^2$ are soft mass squareds of $X$ and $Y$, and $A_{\lambda_X}$ is the $A$-parameter associated with the coupling $\lambda_X$.
Since $X$ and $Y$ are gauge singlets, the direct gauge-mediation contributions to their soft parameters are absent.
However, the Yukawa coupling of $X$ to extra vector-like multiplets $\Psi$, $\bar{\Psi}$ (see \eq{W}) generates soft mass terms at 1-loop level.

The renormalization group equation of $m_X^2$ below the messenger scale is 
\be \label{XmasssqRGE}
\frac{d m_X^2}{d \ln Q} 
= - \frac{1}{8 \pi^2} N_\Psi \sum_i |\lambda_{\Psi_i}|^2 \left( m_X^2 + m_{\Psi_i}^2 + m_{\bar{\Psi}_i}^2 + |A_{\lambda_{\Psi_i}}|^2 \right)
\ee
where $N_\Psi$ is the number of vector-like $\Psi $, $\bar{\Psi}$ pairs, $m_{\Psi_i}^2$ is the soft mass squared of $\Psi_i$(the $i$-th component of $\Psi$) and $A_{\lambda_{\Psi_i}}$ is the $A$-parameter associated with $\lambda_{\Psi_i}$.
Note that $m_X^2$ and $|A_{\lambda_{\Psi_i}}|^2$ in the right-hand side of \eq{XmasssqRGE} are negligibly small during the most part of running from intermediate to weak scale, hence we can ignore their contributions.
In minimal gauge-mediation scenario \cite{Giudice:1998bp}, we obtain the scalar soft masses for vector-like pairs,
\be 
m_{\Psi_i}^2 = m_{\bar{\Psi}_i}^2 = \frac{2}{N_\mathrm{m}} \sum_a C_a(\Psi_i) M_a^2
\ee
where $C_a(\Psi_i)$ is the quadratic Casimir group theory invariants for the superfield $\Psi_i$ for gauge group $G_a$ and $N_m=2\sum_i l_i$ with $l_i$ being the index of the representation of $\Psi_i$.
Thus, we find
\be \label{softmXsq}
m_X^2(X) 
\simeq \frac{1}{2 \pi^2} \frac{N_\Psi}{N_\mathrm{m}} \sum_{i, \, a} |\lambda_{\Psi_i}|^2 C_a(\Psi_i) M_a^2 \ln \left( \frac{M}{|\lambda_{\Psi_i} X|} \right). 
\ee
Meanwhile, from the wave function renormalization of $X$, we find 
\be \label{softAX}
A_{\lambda_X} \simeq - \frac{3}{\pi^2} N_\Psi 
\sum_{i, \, a} C_a(\Psi_i) |\lambda_{\Psi_i}|^2 \left( \frac{\alpha_a}{4 \pi} \right) \ln^2 \left( \frac{M}{|\lambda_{\Psi_i} X|} \right) M_a  
\ee
with $\alpha_a = g_a^2 / \left( 4 \pi \right)$. 
Thus, from \eqs{softmXsq}{softAX}, $m_{X}(X) \gg A_{\lambda_X}(X)$.

Denoting the PQ fields as 
\bea
X &=& \left( X_0 + \frac{x}{\sqrt{2}} \right) \exp \left[ i \frac{a_X}{\sqrt{2} X_0} \right]\,,
\\
Y &=& \left( Y_0 + \frac{y}{\sqrt{2}} \right) \exp \left[ i \frac{a_Y}{\sqrt{2} Y_0} \right]\,,
\eea
we find the physical light and heavy axion states,
\bea
a &=& q_X \frac{X_0}{f_a} a_X + q_Y \frac{Y_0}{f_a} a_Y
\\
a' &=& q_Y \frac{Y_0}{f_a} a_X - q_X \frac{X_0}{f_a} a_Y
\eea
where $f_a = \sqrt{(q_X X_0)^2 + (q_Y Y_0)^2}$.
The mass of the heavy axion $a'$ is 
\be \label{aprimemass}
m_{a'}^2 = \frac{\partial^2 V}{\partial a'^2} = \frac{1}{3} A_{\lambda_X} \frac{\lambda_X f_a^2}{M_\mathrm{GUT}} \frac{X_0}{Y_0} =  \frac{\lambda_X f_a^2}{M_\mathrm{GUT}} \frac{\lambda_X X_0^2}{M_\mathrm{GUT}} \simeq 3 m_{X}^2
\ee 
where we have used $f_a \simeq X_0$.

The elements of the flaton mass matrix are 
\bea
\mathcal{M}_{yy}^2 &=& m_{Y}^2 + \left| \frac{\lambda_X X_0^2}{M_\mathrm{GUT}} \right|^2 \simeq 3 m_{X}^2\,,
\\
\mathcal{M}_{xy}^2 &=& \left| \frac{\lambda_X X_0^2}{M_\mathrm{GUT}} \right| \left( 4 \left| \frac{\lambda_X X_0^2}{M_\mathrm{GUT}} \right| \frac{Y_0}{X_0} - A_{\lambda_X} \right) \simeq \frac{\sqrt{3}}{3} A_{\lambda_X} m_{X}\,,
\\
\mathcal{M}_{xx}^2 &=& \left| \frac{\lambda_X X_0^2}{M_\mathrm{GUT}} \right|^2 \left[ \frac{4}{3} + 4 \left( \frac{Y_0}{X_0} \right)^2 - A_{\lambda_X} \left| \frac{\lambda_X X_0^2}{M_\mathrm{GUT}} \right|^{-1} \left( \frac{Y_0}{X_0} \right) \right] \simeq 4 m_{X}^2\,.
\eea
Due to a small mixing between the flatons, $x$ and $y$, the flaton mass spectra are approximately
\bea
m_{f_1}^2 &\simeq& 3 m_{X}^2(X_0) - \frac{1}{3} A_{\lambda_X}^2\,,
\\
m_{f_2}^2 &\simeq& 4 m_{X}^2(X_0) + \frac{1}{3} A_{\lambda_X}^2\,.
\eea
In the basis of mass eigenstates, $x$ and $y$ are expressed as
\bea
x &=& - \sin \alpha f_1 + \cos \alpha f_2\,,
\\
y &=& \cos \alpha f_1 + \sin \alpha f_2
\eea
where
\bea \label{sin2alpha}
\sin (2 \alpha) &=& \frac{2 \mathcal{M}_{xy}^2}{m_{f_2}^2 - m_{f_1}^2} \simeq \frac{2 \sqrt{3}}{3} \frac{A_{\lambda_X}}{m_{X}}\,,
\\ \label{cos2alpha}
\cos (2 \alpha) &=& \frac{\mathcal{M}_{xx}^2 - \mathcal{M}_{yy}^2}{m_{f_2}^2 - m_{f_1}^2} \simeq 1 - \frac{2}{3} \left( \frac{A_{\lambda_X}}{m_{X}} \right)^2\,.
\eea
Since $m_X \gg A_{\lambda_X}$, we find $x \sim f_2$ and $y \sim f_1$.

The mass matrix of the flatinos has the following nonzero elements,
\bea
\mathcal{M}_\mathrm{\tilde{x} \tilde{x}} &=& 2 \frac{\lambda_X X_0^2}{M_\mathrm{GUT}} \frac{Y_0}{X_0} \simeq \frac{2}{3} A_{\lambda_X}\,,
\\
\mathcal{M}_\mathrm{\tilde{x} \tilde{y}} &=& \frac{\lambda_X X_0^2}{M_\mathrm{GUT}} \simeq \sqrt{3} m_{X}\,.
\eea
whose eigenvalues are
\be
m_{\tilde{f}_{1,2}} \simeq \frac{1}{3} A_{\lambda_X} \mp \sqrt{3} m_{X}\,. \label{flatinomass}
\ee
In the flavor basis, the eigenstates are expressed as
\bea
\tilde{f}_1 &\simeq& \frac{1}{\sqrt{2}} \left( - \tilde{x} + \tilde{y} \right)\,,
\\
\tilde{f}_2 &\simeq& \frac{1}{\sqrt{2}} \left( \tilde{x} + \tilde{y} \right) \,.
\eea
Note that particles in the flaton sector have masses of order $m_{X}$, except the light axion.

\section{Jordan frame supergravity}\label{sec:A}

The Jordan-frame action \cite{jsugra,hmlee} is
\be
S_J=\int d^4 x \sqrt{-g_J}\Big(-\frac{1}{6}\Omega R-\Omega_{i{\bar j}}D_\mu X^i D^\mu {\bar X}^{\bar j}+\Omega b^2_\mu-V_J  \Big)
\ee
where the auxiliary vector field $b_\mu$ take the form,
$b_\mu = -\frac{i}{2\Omega}\Big(D_\mu X^i\partial_i\Omega - D_\mu {\bar X}^{\bar i}\partial_{\bar i}\Omega\Big)$
and the frame function is related to the K\"ahler potential as
$\Omega = - 3 e^{-K/3}$. 
Here the covariant derivatives for scalar fields $X^i$ are given by $D_\mu X^i=\partial_\mu X^i+iA^a_\mu \eta^i_a$.

In order to get the canonical scalar kinetic terms in the Jordan frame, we need $\Omega_{i{\bar j}}=\delta_{i{\bar j}}$ and $b_\mu=0$. The most general frame function for giving $\Omega_{i{\bar j}}=\delta_{i{\bar j}}$ is the following \cite{jsugra,hmlee},
\be
\Omega= - 3 + \delta_{i{\bar j}} X^i {\bar X}^{\bar j} -\frac{3}{2}(F(X)+{\rm h.c.}).\label{framefunc}
\ee
When $F=0$, the non-minimal coupling of the scalar fields are fixed as ${\cal L}=-\sqrt{-g}\,\sum_i\xi_i|X_i|^2R$ with $\xi_i=\frac{1}{6}$ so the scalar fields are conformally coupled to gravity. 
However, by choosing an appropriate holomorphic function $F$, we can break the conformal symmetry explicitly and include the nontrivial non-minimal coupling to gravity. 

Then, from the relation (\ref{framefunc}), the corresponding K\"ahler potential takes the following form,
\be
K= - 3\ln \Big(1-\frac{1}{3}\delta_{i{\bar j}} X^i {\bar X}^{\bar j} +\frac{1}{2}(F(X)+{\rm h.c.})\Big).\label{kahler}
\ee  
Performing a Weyl transformation of the metric with $g^E_{\mu\nu}=(-\Omega/3)g^J_{\mu\nu}$, we obtain the standard Einstein-frame action as
\be
S_E=\int d^4 x \sqrt{-g_E}\Big(\frac{1}{2} R -K_{i{\bar j}}D_\mu X^i D^\mu {\bar X}^{\bar j}-V_E\Big).
\ee
Here the Einstein-frame scalar potential is related to the Jordan-frame one  and is given in terms of the K\"ahler potential $K$, the superpotential $W$ and the gauge kinetic function $f_{ab}$ by
\be
V_E = \frac{9}{\Omega^2}V_J=V_F + V_D \label{spot}
\ee
where
\bea
V_F &=& e^K \Big(K^{i{\bar j}}(D_iW)(D_{\bar j} W^\dagger)-3|W|^2\Big),  \label{fterm}\\
V_D &=& \frac{1}{2} {\rm Re}f^{-1}_{ab} \Big(\eta^i_a\partial_i K\Big)\Big(\eta^i_b\partial_i K\Big).\label{dterm}
\eea

Taking the non-minimal coupling and the superpotential to be
\bea
F(X)&=&\xi_{i j}X^i X^j, \\
W(X)&=&\lambda_{ijk} X^i X^j X^k,
\eea
we obtain the Jordan-frame potential in a simplified form \cite{ferrara},
\be
V_J=\delta^{i{\bar j}}W_i{\bar W}_{\bar j}-\frac{3\Big|\delta^{i{\bar j}}{\bar \xi}_{{\bar j}{\bar k}}{\bar X}^{\bar k}W_i \Big|^2}{1-\frac{1}{2}(\xi_{ij} X^i X^j+{\rm h.c.})+3\delta^{i{\bar j}}\xi_{ij}{\bar\xi}_{{\bar j}{\bar k}}X^j{\bar X}^{\bar k} }.
\ee
In the text, higher order terms are added in the frame function for the stability of non-inflaton fields so that the kinetic terms in Jordan frame become non-canonical.
Then, the Jordan-frame potential is not of the above form any more but it has the corrections coming from those higher order terms as shown in Ref.~\cite{hmlee,ferrara}.

In our model, the minimal frame function and the superpotential relevant for inflation are the following,
\bea
\Omega&=&-3+|Y|^2+\sum_{i=1}^3 |N_i|^2-\frac{3}{2}\Big(\sum_{i=1}^3 \xi_i N_i N_i+{\rm h.c.}\Big), \label{framef}\\
W&=&\frac{1}{2}\sum_{i=1}^3\lambda_{Yi} Y N_i N_i .\label{superp1}
\eea
Then, we find that the Jordan-frame potential is given by
{\small
\bea
V_J=\frac{1}{4}\Big|\sum_{i=1}^3 \lambda_{Yi} N^2_i\Big|^2+|Y|^2 \Big(\sum_{i=1}^3|\lambda_{Yi} N_i|^2\Big)-\frac{3 |Y|^2\Big|\sum_{i=1}^3\lambda_{Yi}\xi^\dagger_i |N_i|^2\Big|^2}{1+\frac{1}{2}\sum_{j=1}^3\Big[6 |\xi_j N_j|^2-({\xi_j N^2_j}+{\rm h.c.})\Big]}. \label{jpot}
\eea
}
For $N_2=N_3=0$ and $N_1\neq 0$, the Jordan-frame potential (\ref{jpot}) becomes
\be
V_J=\frac{1}{4}|\lambda_{Y1}|^2 |N_1|^4+|\lambda_{Y1}|^2 |Y|^2 |N_1|^2-\frac{3|\lambda_{Y1}\xi_1|^2 |Y|^2|N_1|^2}{1+\frac{1}{2}\Big[6 |\xi_1|^2|N_1|^2-({\xi_1 N^2_1}+{\rm h.c.})\Big]}.
\ee
In this case, for $\xi_1|N_1|^2\gg 1$, the potential becomes
\be
V_J\simeq \frac{1}{4}|\lambda_{Y1}|^2 |N_1|^4-\frac{|\lambda_{Y1}|^2|Y|^2|N_1|^2(\xi_1 N^2_1+{\rm h.c.})}{6 |\xi_1|^2|N_1|^2-({\xi_1 N^2_1}+{\rm h.c.})}.\label{jpot2}
\ee
Then, for $|\xi_1|\gg 1$, the flaton $Y$ has the tachyonic instability as follows,
\be
V_J\simeq \frac{1}{4}|\lambda_{Y1}|^2 |N_1|^4-\frac{|\lambda_{Y1}|^2}{6|\xi_1|^2}|Y|^2(\xi_1N^2_1+\xi^\dagger_1 {\overline N}^2_1).
\ee
The tachyonic instability remains even for a smaller $|\xi_1|$, satisfying $|\xi_1|>\frac{1}{3}$, which is needed for a positive effective Planck mass in Jordan frame.
This instability arises due to the sequestered form of the frame function (\ref{framef}), which corresponds to the K\"ahler potential of no-scale type.
Since the large sneutrino VEV breaks SUSY by the F-term of the flaton $Y$, 
we need to add a higher order term, $-\gamma |Y|^4$, in the frame function (\ref{framef}) to generate a positive soft scalar mass for $Y$ during inflation \cite{hmlee}.

From eq.~(\ref{jpot}) with $Y=0$ and $N_1\neq 0$, we also obtain the following effective tachyonic mass terms for the non-inflaton sneutrinos,
\be
V_{J,{\rm numass}}=\frac{1}{4}\lambda^\dagger_{Y1}{\overline N}^2_1(\lambda_{Y2} N^2_2+\lambda_{Y3} N^2_3)+{\rm h.c.}
\ee
Thus, the direction satisfying $N_2=N_3=0$ would be unstable. This instability is cured by adding additional higher order terms, $-\delta_2|Y|^2|N_2|^2$ and $-\delta_3|Y|^2|N_3|^2$, in the frame function (\ref{framef}). With these higher order terms, the nonzero F-term SUSY breaking of the flaton $Y$ is transmitted to the non-inflaton sneutrinos such that their positive soft scalar masses are generated.
Therefore, the above discussion brings us to the final form of the frame function 
\be \label{framefull0}
\Omega=-3+|\Phi_j|^2+|Y|^2\Big(1-\gamma|Y|^2 - \sum_{i \neq 1} \delta_i |N_i|^2\Big)+\sum_{i=1}^3\Big[|N_i|^2-\frac{3}{2}( \xi_i N_i N_i+{\rm h.c.})\Big]
\ee
where $\Phi_i$ are all the chiral superfields in the model, except the flaton $Y$
and the sneutrinos $N_i$.

\section{Number of efoldings with thermal inflation}
\label{sec:efoldings}

In the presence of late-time thermal inflation, the total entropy is conserved once the universe is completely reheated after thermal inflation.
The total entropy at the time of the flaton decay is given by 
\be
S_\mathrm{d} \equiv R_\mathrm{d}^3 s_d
\ee
where 
\be
R_\mathrm{d} = R_* 
\left( \frac{a_e}{a_*} \right)
\left( \frac{a_\mathrm{t}}{a_e} \right)
\left( \frac{a_\mathrm{b}}{a_\mathrm{t}} \right)
\left( \frac{a_\mathrm{c}}{a_\mathrm{b}} \right)
\left( \frac{a_\mathrm{d}}{a_\mathrm{c}} \right)
\ee
with
\bea
\left( \frac{a_e}{a_*} \right) &=& e^{N_e}
\\
\left( \frac{a_\mathrm{t}}{a_e} \right) &=& \left( \sqrt{3} \xi_1 \right)^{2/3}
\\
\left( \frac{a_\mathrm{b}}{a_\mathrm{t}} \right) &\simeq& \left( \frac{\pi^2}{30} g_*(T_\mathrm{b}) \right)^{-1/4} \frac{V^{1/4}(\varphi_\mathrm{t})}{T_\mathrm{b}}
\\
\left( \frac{a_\mathrm{c}}{a_\mathrm{b}} \right) &=& e^{N_{TI}}
\\
\left( \frac{a_\mathrm{d}}{a_\mathrm{c}} \right) &\simeq& \left( \frac{\pi^2}{30} g_*(T_\mathrm{d}) \right)^{-1/3} \left( \frac{V_0}{T_\mathrm{d}^4} \right)^{1/3}.
\eea
where the subscripts of scale factor $a$ represent respectively the epochs of  
\begin{itemize}
\item *: Horizon exit of our cosmological scale during inflation
\item e: End of inflation
\item t: Phase transition of inflaton from matter to radiation
\item b: Beginning of thermal inflation
\item c: End of thermal inflation
\item d: Decay of flaton (reheating after thermal inflation)
\end{itemize}
and $\varphi$ is the inflaton field with the potential $V(\varphi)$ while $V_0$ is the vacuum energy during thermal inflation.
Thus, using $R_*=\frac{1}{H(\varphi_*)}$, we obtain
\be
S_\mathrm{d}
\simeq
\frac{1.7 \times 10^3}{H^3(\varphi_*)} \,e^{3 N_e}\, \frac{V^{3/4}(\varphi_\mathrm{t})}{T_\mathrm{b}^3}\, e^{3 N_{TI}}\, \frac{V_0}{T_\mathrm{d}}
\ee
where we have used $s_\mathrm{d} = \left( 2 \pi^2 / 45 \right) g_{s*}(T_\mathrm{d}) T_\mathrm{d}^3$ and $g_*(T_\mathrm{b})=g_*(T_\mathrm{d})=g_{s*}=200$.
Therefore, from $S_{\rm d}=S_0$, the number of efoldings necessary for the primordial inflation is given by
\be \label{R0-e-foldings}
N_e(R_0) \simeq \frac{1}{3} \ln S_0 -2.5 - N_\mathrm{TI} - \ln \left( \frac{V^{1/4}(\varphi_\mathrm{t})}{T_\mathrm{b}} \right) - \frac{1}{3} \ln \left( \frac{V_0}{H^3(\varphi_*) T_\mathrm{d}} \right)
\ee
where $R_0 \sim 3000\, \mathrm{Mpc}$ and $S_0 \sim 10^{88}$ are the present Hubble radius and the total entropy in the Hubble patch.
From \eq{N-exit}, we find
\be
H(\varphi_*) = \frac{V^{1/2}(\varphi_*)}{\sqrt{3} M_\planck} \simeq 6.8 \times 10^{12} \GeV \left( \frac{\lambda_{Y1}}{10^{-3}} \right) \left( \frac{10^2}{\xi_1} \right)
\ee
and 
\be
V(\varphi_\mathrm{t}) \simeq \frac{1}{4} \lambda_{Y1}^2 \varphi_{ \mathrm{t}}^4
\ee
where $\varphi_{\mathrm{t}} \simeq \sqrt{\frac{2}{3\xi_1}}$ is the inflaton field value when inflaton starts to behave as a radiation  \cite{reheating}.
Therefore, taking 
$H(\varphi_*) = 10^{13} \GeV$
and $V^{1/4}(\varphi_\mathrm{t}) = 10^{14} \GeV$
with $N_\mathrm{TI} = 8$, $T_\mathrm{b} = 10^6 \GeV$ and $T_\mathrm{d} = 1 \TeV$, we find $N_e(R_0) \simeq 52$.

\section{Critical temperatures}\label{sec:ctemp}

The critical temperature, at which a field $\varphi$ becomes unstable around the origin, is given by
\be
T_\varphi = \frac{m_\varphi}{\beta_\varphi}
\ee
where $m_\varphi \equiv \sqrt{|m^2_\varphi(0)|}$ with $m^2_\varphi(0)$ being the curvature of the potential along $\varphi$ around the origin, and $\beta_\varphi$ is given by \cite{Comelli:1996vm}
\be
\beta_\varphi^2 = \frac{1}{8} \left( \sum_{ij} |\lambda_{\varphi ij}|^2 + 4 \sum_a C_a(\varphi) g_a^2 \right)
\ee
In case of the $LH_u$ flat-direction, one finds 
\bea
m^2_{LH_u} &\equiv& - \frac{1}{2} \left( m^2_L + m^2_{H_u} \right)\,, \label{lhumass}
\\
\beta_{LH_u}^2 &\equiv& \frac{1}{2} \left( \beta_L^2 + \beta_{H_u}^2 \right) = \frac{1}{8} \left( 3 |\lambda_t|^2 + |\lambda_\tau|^2 + 3 g_2^2 + \frac{3}{5} g_1^2 \right)\,.
\eea 
The RGE of $m^2_{LH_u}$ is \cite{Martin:1997ns}
\bea
\frac{d m^2_{LH_u}}{d \ln Q} 
\simeq - \frac{1}{16 \pi^2} \left[ 3 |\lambda_t|^2 \left( m^2_{H_u} + m^2_Q + m^2_{\bar{u}} + |A_t|^2 \right) - 6 g_2^2 |M_2|^2 - \frac{6}{5} g_1^2 |M_1|^2 \right]\,.
\eea
In minimal gauge mediation, gaugino masses and soft scalar masses are given by \cite{Giudice:1998bp},
\bea
M_a &=& N_m \frac{\alpha_a}{4 \pi} \frac{F}{M}\,,
\\
m_i^2 &=& 2 N_m \sum_a C_a \left( \frac{\alpha_a}{4 \pi} \right)^2 \left( \frac{F}{M} \right)^2\,.
\eea
Hence, the scalar soft masses at the messenger scale are
\bea
m^2_{H_u} &\simeq& \frac{3}{2 N_\mathrm{m}} M_2^2\,,
\\
m^2_Q &\simeq& \frac{8}{3 N_\mathrm{m}} M_3^2 = m^2_{\bar{u}}\,.
\eea
So,  using $\alpha_3 = 2 \alpha_2$, we get
\be
\frac{m^2_{H_u}}{m^2_Q} \simeq \frac{9}{16} \left( \frac{\alpha_2}{\alpha_3} \right)^2 = \frac{9}{64}\,.
\ee
At the messenger scale, the $A$-term contribution is negligible and the contribution of $m^2_{H_u}$ is largely cancelled by the contributions of gauge interactions, so we ignore those contributions in RGE of $m^2_{LH_u}$.
Then, one finds
\bea
m^2_{LH_u}(0) 
\label{msqLHu}
\sim \frac{1}{\pi^2 N_\mathrm{m}} |\lambda_t|^2 M_3^2 \ln \frac{M}{m_\mathrm{soft}}\,.
\eea

Meanwhile, in case of the flaton $X$,
from \eq{softmXsq}, we find
\be \label{softmXsq0}
m_X^2(0) \sim \frac{1}{2 \pi^2} \frac{N_\Psi}{N_\mathrm{m}} \sum_a N_a |\lambda_{\Psi_i}|^2 C_a(\Psi_i) M_a^2 \ln \left( \frac{M}{m_\mathrm{soft}} \right)
\ee
and 
\be
\beta_X^2 = \frac{1}{4} N_{\Psi} \sum_i |\lambda_{\Psi_i}|^2 \,.
\ee
Therefore, we find the ratio of the critical temperature for thermal inflation ($T_c$) to the one for destabilizing the AD field ($T_{LH_u}$) as
\bea \label{rTc}
\frac{T_\mathrm{c}}{T_{LH_u}} 
&=& \frac{m_X(0)}{m_{LH_u}(0)} \left( \frac{\beta_{LH_u}}{\beta_X} \right)
\nonumber \\
&\lesssim& \sum_a \sqrt{\frac{N_\Psi N_a C_a(\Psi_i)}{2}} \left( \frac{\lambda_{\Psi_i}}{\lambda_t} \right) \left( \frac{M_a}{M_3} \right) \left( \frac{3 |\lambda_t|^2 + 3 g_2^2 + (3/5) g_1^2	}{ 2 N_\Psi \sum_i |\lambda_{\Psi_i}|^2} \right)^{1/2}
\nonumber \\
&\simeq& \sqrt{ \frac{7}{6}} \sqrt{\frac{3}{5}} \, \left( 1 + \frac{g_2^2 + g_1^2/5}{\lambda_t^2} \right)^{1/2}
\eea
where in the last line we have assumed $M_1:M_2:M_3 = 1:2:6$.
From \eq{rTc}, it is easy to see that $T_{LH_u} > T_\mathrm{c}$ is always satisfied.
Therefore, the $LH_u$ flat direction will be destabilized earlier than the flaton $X$.

\end{document}